\documentclass[aps,prl,twocolumn,superscriptaddress,amssymb,amsmath,nofootinbib]{revtex4}

\usepackage{color}
\usepackage{epsfig}
\usepackage{graphicx}
\usepackage{float}
\usepackage{epstopdf}
\usepackage{hyperref} 
\usepackage[usenames,dvipsnames]{xcolor}
\usepackage{multirow}
\usepackage[normalem]{ulem}
\usepackage{slashed}
\usepackage{feynmf}

\begin{document}

\preprint{}

\title{The Ellipse of Muon Dipole Moments }

\author{Radovan Dermisek}
\email[]{dermisek@indiana.edu}
\affiliation{Department of Physics, Indiana University, Bloomington, IN, 47405, USA}

\author{Keith Hermanek}
\email[]{khermane@iu.edu}
\affiliation{Department of Physics, Indiana University, Bloomington, IN, 47405, USA}

\author{Navin McGinnis}
\email[]{nmcginnis@triumf.ca}
\affiliation{TRIUMF, 4004 Westbrook Mall, Vancouver, BC, Canada V6T 2A3}

\author{Sangsik Yoon}
\email[]{yoon12@iu.edu}
\affiliation{Department of Physics, Indiana University, Bloomington, IN, 47405, USA}

\date{\today}

\begin{abstract} 
We show that any new interaction resulting in a chirally-enhanced contribution to the muon magnetic moment necessarily modifies 
the decay rate of the Higgs boson to muon pairs or generates the muon electric dipole moment. These three observables are highly correlated, and near future measurements of $h\to \mu^+\mu^-$ will carve an ellipse in the plane of dipole moments for any such model. Together with the future  measurements of the electric dipole moment many models able to explain the muon $g-2$ anomaly can be efficiently tested.
\end{abstract}


\maketitle


\noindent
{\it Introduction.}
Besides the mass, electric charge, and other quantum numbers, the calculable electric and magnetic dipole moments are  among the basic attributes of elementary particles. 
Together with the Yukawa coupling inferred from the mass, they provide excellent opportunities to test  the standard model (SM) and   probe new physics. 

\begin{figure}[t]
\includegraphics[width=0.71\linewidth]{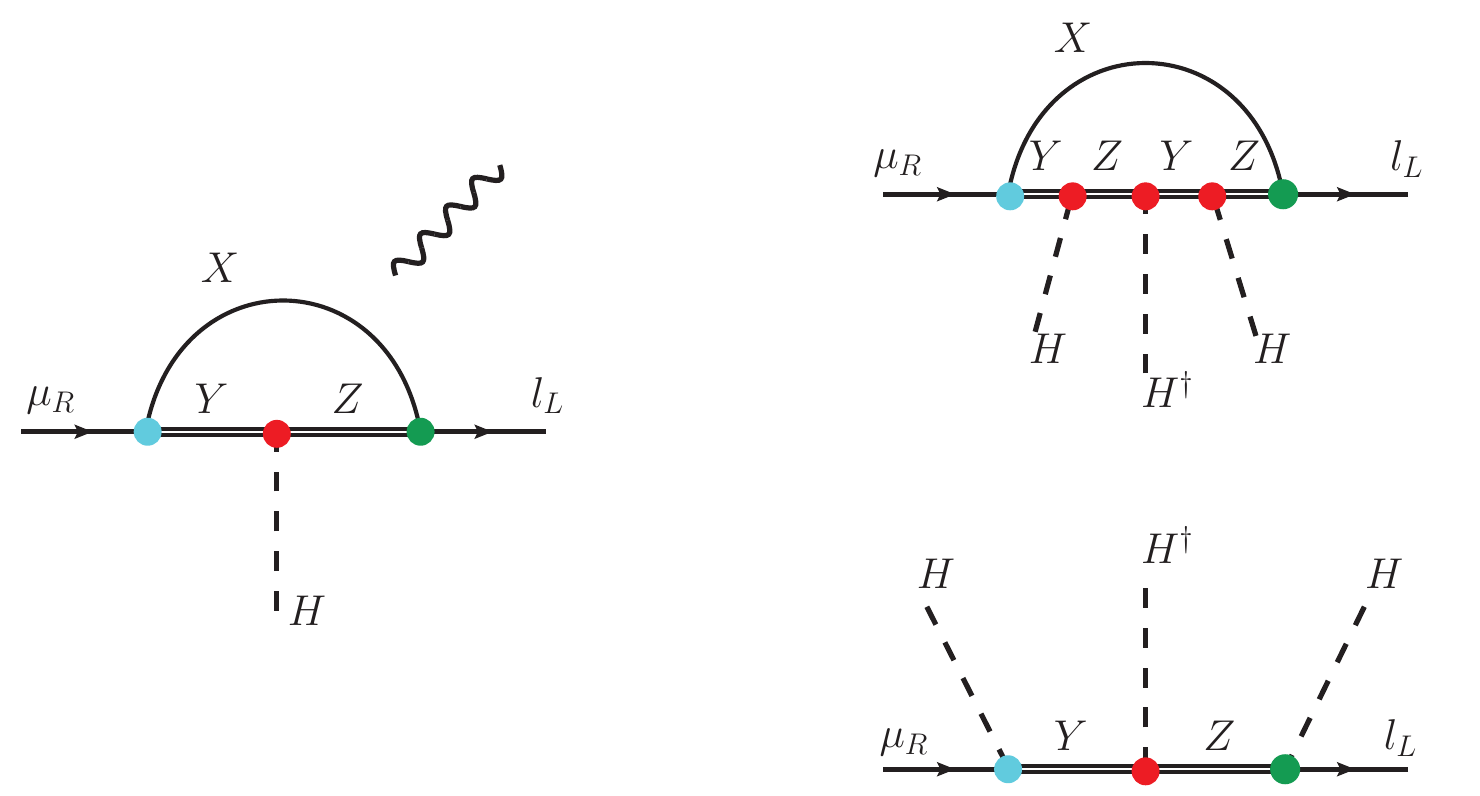}
\caption{A generic diagram with chiral enhancement contributing to muon dipole moments (left), a corresponding diagram contributing to the dimension 6 mass operator at 1-loop level  (right top) and at tree level, if possible, (right bottom).}
\label{fig:diags}
\end{figure}

The 4.2$\sigma$  deviation of the measured value of the muon anomalous magnetic moment from the SM prediction, $\Delta a_\mu   = (2.51 \pm 0.59)\times 10^{-9}$~\cite{Abi:2021gix, Aoyama:2020ynm},  can be comfortably explained even by very heavy new particles as a result of the chiral enhancement from the Higgs coupling to new particles, see Fig.~\ref{fig:diags} (left). With order one couplings, the scale of new physics up to $\sim 10$ TeV  is expected, and it further extends to $\gtrsim 50$ TeV for couplings close to the perturbativity limit~\cite{Dermisek:2020cod,Capdevilla:2021rwo,Dermisek:2021ajd,Allwicher:2021jkr}. Such heavy particles are far beyond the reach of the Large Hadron Collider (LHC) and currently envisioned future experiments.

In this Letter we show that any new interaction resulting in a chirally enhanced contribution to the muon magnetic moment necessarily modifies the muon Yukawa coupling, and thus the decay of the Higgs boson to muon pairs, or, if $h\to \mu^+\mu^-$ is not modified, the muon electric dipole moment ($\mu$EDM) of certain size must be generated (with one exception noted later). These three observables are highly correlated, and near future measurements of $h\to \mu^+\mu^-$ will carve an ellipse in the plane of dipole moments for any such model. Together with the improved  measurement of the electric dipole moment many models able to explain $\Delta a_\mu$ can be efficiently tested.
Furthermore, in some scenarios the heaviest possible spectrum will be tested the most efficiently. 

The main results can be intuitively understood from Fig.~\ref{fig:diags}. No matter what the quantum numbers of $X$, $Y$, and $Z$ particles are, as long as they can form the diagram on the left between $SU(2)$ doublet, $l_{L}$, and singlet, $\mu_{R}$, leptons,  the photon can be removed and the $Y$-$Z$-$H$ coupling and its conjugate can be used again to generate the diagram on the right top. This effectively generates the dimension 6 mass operator, $\bar{l}_{L}\mu_{R}H\left(H^{\dagger}H\right)$. In addition, for models where $X$ is a scalar participating in electroweak symmetry breaking, for example, the SM Higgs boson, the same operator could be generated at tree level as in the diagram on the right bottom. We refer to these cases hereafter as loop models and tree models, respectively. The tree models have been  studied in connection with $\Delta a_{\mu}$ in~\cite{Kannike:2011ng,Dermisek:2013gta,Dermisek:2014cia,Poh:2017tfo,Crivellin:2018qmi,Dermisek:2020cod,Dermisek:2021ajd}, whereas examples of loop models,~\cite{Moroi:1995yh,Huang:2001zx,Cheung:2009fc,Endo:2013lva,Freitas:2014pua,Thalapillil:2014kya,Omura:2015nja,Calibbi:2018rzv,Crivellin:2018qmi,Crivellin:2020tsz,Babu:2020hun,Capdevilla:2021rwo,Crivellin:2021rbq,Babu:2021jnu,Bigaran:2021kmn,MuonCollider:2022xlm}, include scenarios with familiar particles in the loop: superpartners, top quark, or the $\tau$ lepton; and particles solely introduced to explain $\Delta a_\mu$. The generated operator contributes differently to the muon mass and Yukawa coupling as a result of different combinatorial factors. This necessarily modifies the rate for $h\to \mu^+\mu^-$, unless the modified Yukawa coupling has the same magnitude as that in the SM, which is possible with complex couplings that in turn predict  a certain value of $\mu$EDM.

Possible correlations between $\Delta a_\mu$ and $h\to \mu^+\mu^-$ were pointed out before~\cite{Kannike:2011ng,Dermisek:2013gta,Dermisek:2014cia,Thalapillil:2014kya,Poh:2017tfo,Crivellin:2020tsz,Babu:2020hun,Dermisek:2020cod,Dermisek:2021ajd,Crivellin:2021rbq}. Similarly $\mu$EDM was also studied in connection with $\Delta a_{\mu}$ but only as a possible effect if couplings are complex~\cite{Cheung:2009fc,Crivellin:2018qmi,Babu:2020hun,Bigaran:2021kmn,MuonCollider:2022xlm}. The sharp correlation between all three observables has not been noticed.  As we will see, with complex couplings, predictions for $h\to \mu^+\mu^-$ cannot be made based only on $\Delta a_\mu$. Rather a given $h\to \mu^+\mu^-$ translates into a prediction for $\mu$EDM and vice versa.

\noindent
{\it Effective Lagrangian.} For our discussion, the relevant terms of the effective Lagrangian are
\begin{eqnarray}
\mathcal{L}\supset  &-&y_{\mu}\bar{l}_{L}\mu_{R}H \;-\; C_{\mu H}\bar{l}_{L}\mu_{R}H\left(H^{\dagger}H\right) \nonumber \\
&-& C_{\mu \gamma} \bar{l}_{L}\sigma^{\rho\sigma}\mu_{R} H F_{\rho\sigma} + h.c.,
\label{eq:eff_lagrangian}
\end{eqnarray}
%
where the components of the lepton doublet are  $l_{L}=(\nu_{\mu}, \mu_{L})^{T}$, $\sigma^{\rho\sigma}=\frac{i}{2}[\gamma^{\rho},\gamma^{\sigma}]$, and all the parameters  can be complex. The first term  is the  usual muon Yukawa coupling in the SM. When the Higgs field develops a vacuum expectation value, $H=(0,v+h/\sqrt{2})^T$ with $v=174$ GeV, the dimension 6 operator in the second term generates additional contributions to the muon mass and muon coupling to the Higgs boson, while the dimension 6 operator in the third term corresponds to muon dipole moments.  Defining the muon Yukawa coupling and the electric and magnetic dipole moments in terms of Dirac spinors in the basis where the muon mass, $m_\mu$, is real and positive,
\begin{eqnarray}
\mathcal{L}\supset &&-  m_{\mu} \bar{\mu}\mu - \frac{1}{\sqrt{2}} \left(\lambda^{h}_{\mu\mu}\bar{\mu}P_{R}\mu h + h.c.\right)\nonumber \\
&&\;\;+ \frac{\Delta a_{\mu}e}{4m_{\mu}}\bar{\mu}\sigma^{\rho\sigma}\mu F_{\rho\sigma} - \frac{i}{2}d_{\mu}\bar{\mu}\sigma^{\rho\sigma}\gamma^{5}\mu F_{\rho\sigma},
\label{eq:eff_lagrangian_2}
\end{eqnarray}
%
%
%
we have 
\begin{eqnarray}
m_{\mu}&=&\left(y_{\mu}v + C_{\mu H}v^{3}\right)e^{-i\phi_{m_{\mu}}}, \label{eq:mmu}\\
\lambda_{\mu\mu}^{h}&=&\left(y_{\mu} + 3C_{\mu H}v^{2}\right)e^{-i\phi_{m_{\mu}}}, \label{eq:mmu_lamhmu} \\
\Delta a_{\mu} &=&- \frac{4m_{\mu}v}{e}\textrm{Re}[C_{\mu \gamma}e^{-i\phi_{m_{\mu}}}], \label{eq:mdipole}\\
d_{\mu} &=& 2v\textrm{Im}[C_{\mu \gamma}e^{-i\phi_{m_{\mu}}}],
\label{eq:edipole}
\end{eqnarray}
where $e$ is positive and $\phi_{m_{\mu}}$ is the phase of the rotation required to make the mass term real and positive. All the parameters are real except for $\lambda_{\mu\mu}^{h}$ which can be complex. The factor of 3 in Eq.~(\ref{eq:mmu_lamhmu}), originating from 3 possible ways that one $h$ and two factors of $v$ can be selected from three $H$ in the dimension 6 operator, results in
$\lambda_{\mu\mu}^{h}$ and $m_{\mu}$ not following the expected scaling in the SM and 
\begin{equation}
R_{h\to \mu^+\mu^-} \equiv \frac{BR(h\to \mu^+\mu^-)}{BR(h\to \mu^+\mu^-)_{SM}} = \left(\frac{v}{m_{\mu}}\right)^{2}\big|\lambda_{\mu\mu}^{h}\big|^{2}
\end{equation}
in general deviating from 1.


\noindent
{\it The muon ellipse.} The crucial observation is that the couplings which generate chirally-enhanced contributions to $C_{\mu\gamma}$, also necessarily generate $C_{\mu H}$ with the same phase. Although all the couplings in diagrams in Fig.~\ref{fig:diags} can be complex, the same combination of couplings enter both $C_{\mu\gamma}$ and $C_{\mu H}$ for  tree models, with an additional factor of $\lambda_{YZ} \lambda_{YZ}^*$ for  loop models. Thus, the two Wilson coefficients are related by a {\it real factor}, $k$, defined as
\begin{equation}
C_{\mu H} = \frac{k}{e} C_{\mu\gamma}.
\label{eq:WC_relation}
\end{equation}
This allows us to write $\lambda_{\mu\mu}^{h}$ and thus $R_{h\to \mu^+\mu^-}$ in terms of electric and magnetic dipole moments:
\begin{flalign}
R_{h\to \mu^+\mu^-}=\left(\frac{\Delta a_{\mu}}{2\omega} - 1\right)^{2} + \left(\frac{m_{\mu}d_{\mu}}{e\omega}\right)^{2},
\label{eq:ellipse}
\end{flalign}
where $\omega = m_{\mu}^{2}/kv^{2}$. Note that $\Delta a_{\mu}$ can both increase or decrease $R_{h\to \mu^+\mu^-}$ depending on its sign and the sign of $k$, while $d_{\mu}$ can only increase $R_{h\to \mu^+\mu^-}$.

\begin{figure}[t]
\includegraphics[width=0.65\linewidth]{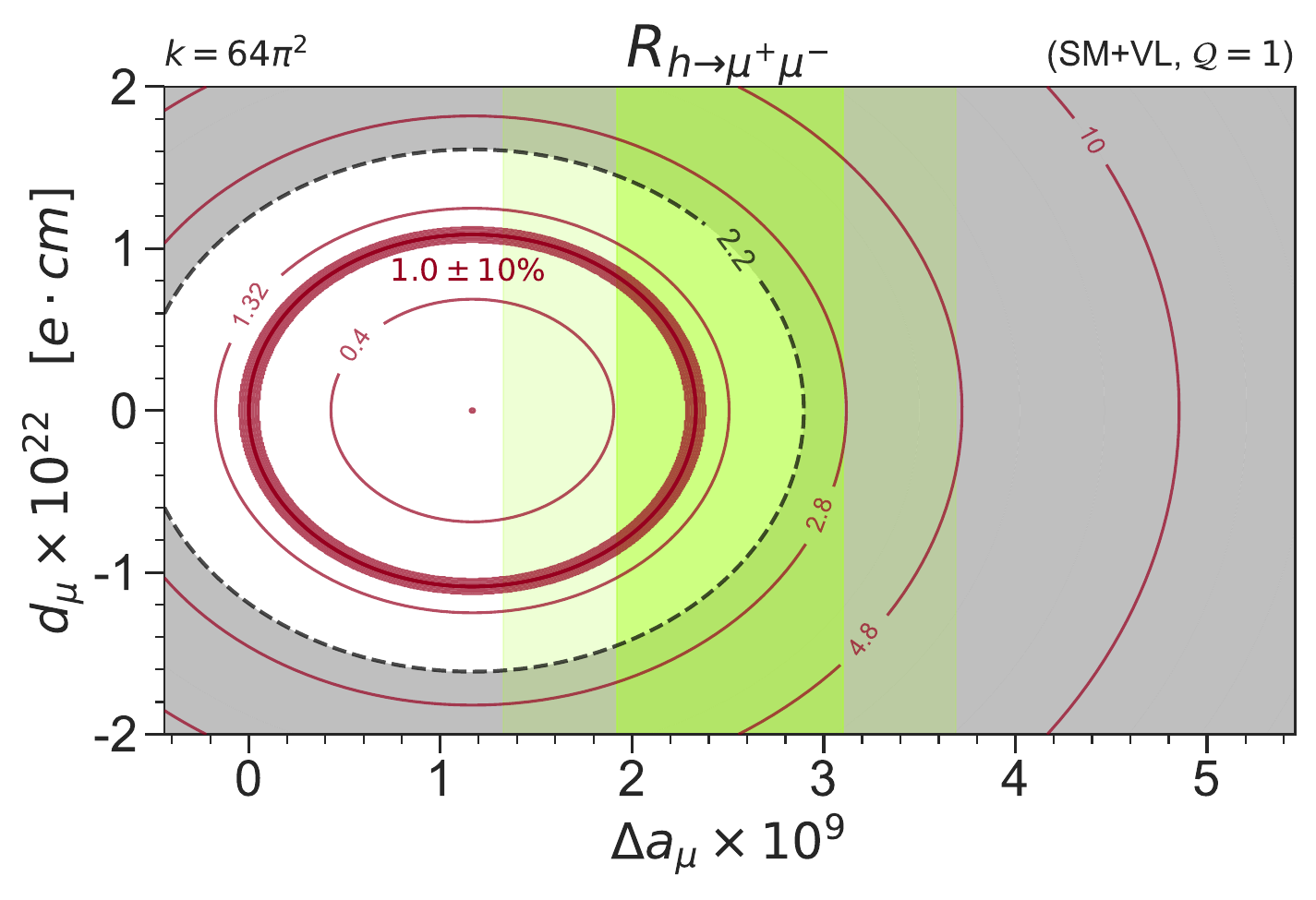} 
\includegraphics[width=0.65\linewidth]{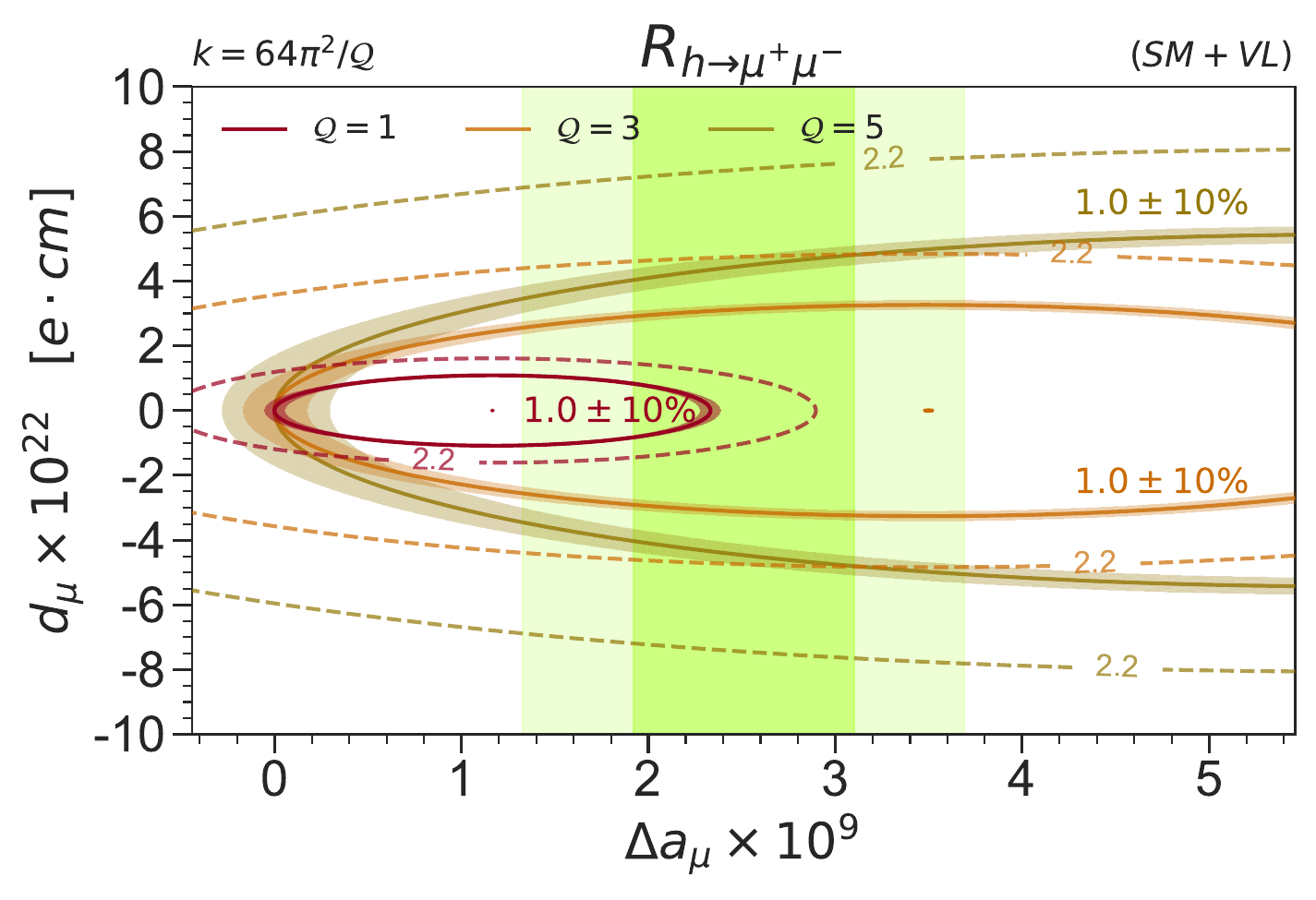} 
\includegraphics[width=0.65\linewidth]{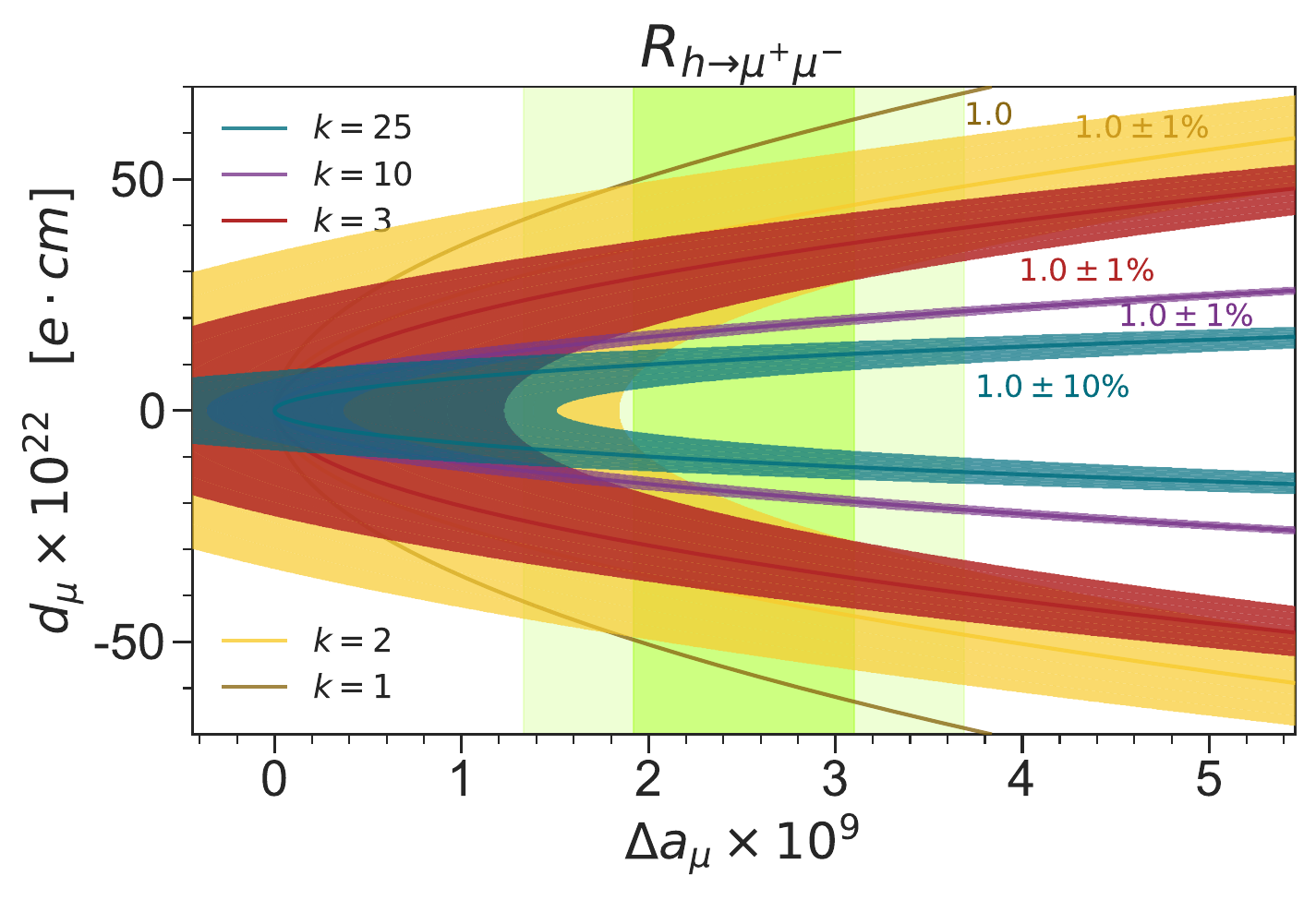} 
\caption{Contours of constant $R_{h\to\mu^+\mu^-}$ in the $\Delta a_{\mu}$ -- $d_{\mu}$ plane in models with $k=64\pi^2$ (e.g., SM+VL, $\mathcal{Q}=1$) (top); $k=64\pi^2, \;64\pi^2/3$, and  $64\pi^2/5$ (e.g., SM+VL, $\mathcal{Q}=1$, 3, and 5) (middle); and for smaller values of $k$ relevant for models with extended Higgs sectors and loop models (bottom). The light and dark green shaded regions show the $\pm 1\sigma$ and $\pm 2\sigma$ ranges of $\Delta a_{\mu}$, respectively. In the top panel, the gray shaded region, where $R_{h\to \mu^+\mu^-}> 2.2$, is already ruled out by measurements of $h\to \mu^{+}\mu^{-}$.}
\label{fig:SM_ellipse}
\end{figure}

Concretely, in five possible extensions of the SM with vectorlike leptons (VL) that can generate  chirally enhanced contributions, $k$ is completely determined by the quantum numbers of new leptons,
\begin{equation}
k=\frac{64\pi^{2}}{\mathcal{Q}},
\label{eq:tree_X_SM}
\end{equation}
where $\mathcal{Q}= 1$  for $X$ and $Y$ leptons in $\mathbf{2}_{-1/2}\oplus\mathbf{1}_{-1}$ or $\mathbf{2}_{-1/2}\oplus\;\mathbf{3}_{0}$ representations of $SU(2)\times U(1)_Y$; $\mathcal{Q}=3$ for $\mathbf{2}_{-3/2}\oplus\mathbf{1}_{-1}$ or $\mathbf{2}_{-3/2}\oplus\mathbf{3}_{-1}$; and $\mathcal{Q}=5$ for $\mathbf{2}_{-1/2}\oplus\mathbf{3}_{-1}$~\cite{Kannike:2011ng}.

In Fig.~\ref{fig:SM_ellipse} (top), we show contours of constant $R_{h\to \mu^+\mu^-}$ in the plane of the muon dipole moments for $k=64\pi^2$ which corresponds, for example, to the SM extended by a vectorlike doublet and singlet leptons whose quantum numbers mirror their respective SM counterparts, i.e., $\mathcal{Q}=1$. The region where $h\to \mu^{+}\mu^{-}$ is found to be within $10\%$ of the SM value, indicating the ultimate LHC precision, is shaded red. The region outside  $R_{h\to \mu^+\mu^-}=2.2$ (shaded gray) is already ruled out by measurements of $h\to \mu^{+}\mu^{-}$~\cite{ATLAS:2020fzp}. This model (range of $k$) is somewhat special as,  in spite of the large contribution from   $C_{\mu H}$ to the muon mass and Yukawa coupling, the SM-like $R_{h\to \mu^+\mu^-}$ and $d_{\mu} = 0$ are consistent with $\Delta a_{\mu}$ within $1\sigma$.\footnote{ This  illustrates the only exception to generating non-zero $\mu$EDM when $R_{h\to \mu^+\mu^-}=1$, advertised in the introduction. It corresponds to the case when $\lambda_{\mu\mu}^{h} = -m_\mu/v$.}
Note, however, that the current central value of $\Delta a_{\mu}$  requires $R_{h\to \mu^+\mu^-} = 1.32$ for $d_{\mu}=0$ and it can be as large as the current upper limit when $|d_{\mu}|\sim 1\times 10^{-22}\, {\rm e \cdot cm}$. 

The situation is dramatically different for the other two scenarios, $\mathcal{Q} = 3$ and 5, with the comparison of all three scenarios shown in  Fig.~\ref{fig:SM_ellipse} (middle). As $k$ decreases,  the center of the ellipse moves to larger values of $\Delta a_{\mu}$. Contours of $R_{\mu}= 1\pm 10\%$ are now consistent with the whole $2\sigma$ range of  $\Delta a_{\mu}$. 
 However, in contrast to the case with $\mathcal{Q}=1$, consistency of $R_{\mu}= 1\pm 10\%$  with $\Delta a_{\mu}$ necessarily implies a nonzero value of $d_{\mu}$. In fact, the consistency sharply requires values of $d_{\mu}\simeq 2.7-3.4\times10^{-22} \,{\rm e\cdot cm}$ ($\mathcal{Q}=3$) and $d_{\mu}\simeq 3.6-5.1\times10^{-22} \,{\rm e\cdot cm}$ ($\mathcal{Q}=5$), which are within the expected sensitivity of  future measurements. Thus, for these scenarios,  the correlation of three observables requires deviations from SM predictions either in $R_{\mu}$ or $d_{\mu}$ that are observable in the near future.

Models where the SM Higgs acts as only a single component of an extended Higgs sector participating in EWSB, such as in a 2HDM, also fall into the class of tree models. However, the mixing in the extended Higgs sector will generically introduce an additional free parameter. In the case of a 2HDM type-II, it is  the ratio of vacuum expectation values of the two Higgs doublets, $\tan\beta$. From the results of Refs.~\cite{Dermisek:2021ajd,Dermisek:2021mhi} we can find that the modification to Eq.~(\ref{eq:tree_X_SM}), assuming a common mass for new leptons much larger than that of heavy Higgses, becomes
\begin{equation}
k=\frac{64\pi^{2}}{\mathcal{Q}(1+\tan^{2}\beta)},
\label{eq:tree_X_2HDM}
\end{equation}
which remains a very good approximation for arbitrary splitting in the spectrum of new leptons and 
Higgs masses comparable to those of new leptons. For low $\tan\beta$ the results are similar as for the SM extensions with VLs discussed above. However, as $\tan\beta$ increases, much smaller $k$ values are possible. Contours of constant $R_{h\to \mu^+\mu^-} = 1$ for a few representative choices of smaller $k$ are shown in  Fig.~\ref{fig:SM_ellipse} (bottom). We also show corresponding $\pm 10\%$ and $\pm 1\%$ regions for cases when the region does not extend all the way to $d_{\mu} = 0$ in the $1\sigma$ range  of  $\Delta a_{\mu}$. For the 2HDM type-II extended with vectorlike leptons with the same quantum numbers as SM leptons, $\mathcal{Q}=1$, the plotted values of $k$ correspond to $ \tan \beta \simeq 5, 8, 14, 18$, and 25; while the 3 cases in the middle plot correspond to $ \tan \beta \simeq 0, 1.4$, and 2 (with the first case not being physical).

 %
%

The loop models with two new fermions and one scalar (FFS) or one new fermion and two scalars (SSF) represent infinite classes of models as the required couplings alone do not completely determine the quantum numbers of new particles. In this case the $k$ factor is directly linked to the coupling responsible for the chiral enhancement, $\lambda_{YZ}$, see Fig.~\ref{fig:diags}. For the FFS models and, in the limit of a common mass of all new particles, we find
\begin{equation}
k=\frac{4}{\mathcal{Q}}|\lambda_{YZ}|^{2}.
\label{eq:loop_Q}
\end{equation}
For SSF models the $Y-Z-H$ coupling $A_{YZ}$ is dimensionful and in the above formula $|\lambda_{YZ}|^2$ should be replaced by $|A_{YZ}|^{2}/M^{2}$ where $M$ is the mass of new particles. The $\mathcal{Q}$ factor is determined by the charges of new particles, and it can be obtained, for example, from the entries in Tables~1 and 2 of Ref.~\cite{Crivellin:2021rbq}. We find that for hypercharge choices $\pm 1/2$ and $\pm 1$ of the scalar in FFS models or fermion in SSF models, $|\mathcal{Q}|$ varies between 1/5 and 6. For small $\mathcal{Q}$ and large $\lambda_{YZ}$,  $k$ can be as large as in the tree  models, the SM+VL or 2HDM type-II at small $ \tan \beta$, while for larger $\mathcal{Q}$ and/or small $\lambda_{YZ}$ the range of $k$ coincides with predictions of 2HDM type-II at  larger $ \tan \beta$. For example, for  $\mathcal{Q}= 1$, the range of $\lambda_{YZ}$ from $0.5$ to $\sqrt{4\pi}$ corresponds to $k = 1$ to $\simeq 50$.

\begin{figure}[t]
\includegraphics[width=0.75\linewidth]{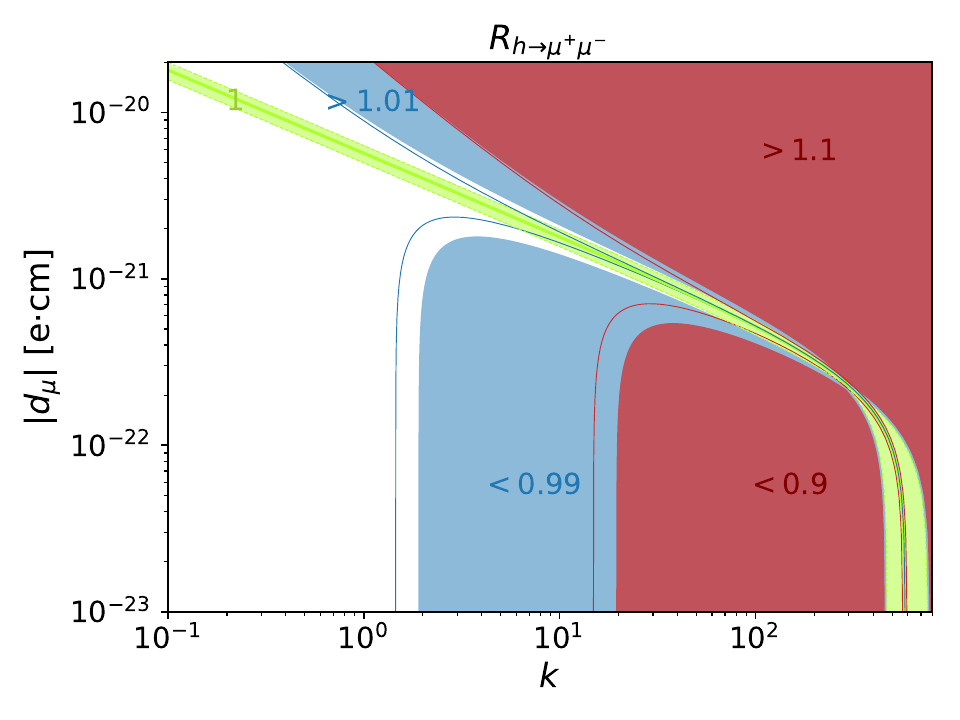} 
\caption{Future exclusion regions in the $k$ -- $|d_{\mu}|$ plane assuming $R_{h\to \mu^+\mu^-}$ is measured to be SM-like, $R_{h\to \mu^+\mu^-} = 1\pm0.1$ (red) and $R_{h\to \mu^+\mu^-} = 1\pm0.01$ (blue) assuming  $\Delta a_{\mu}$ within $1\sigma$ of the measured value. The regions would extend to the solid lines with the same color if the central value of $\Delta a_{\mu}$ was assumed. The green line (shaded region) corresponds to $R_{h\to \mu^+\mu^-} = 1$ assuming   the central value ($1\sigma$ range) of $\Delta a_{\mu}$.}
\label{fig:edm_vs_k}
\end{figure}

From Fig.~\ref{fig:SM_ellipse} we see that as $k$ is decreasing from the values typical for SM+VLs, the consistency of  $\Delta a_{\mu}$ with $R_{h\to \mu^+\mu^-} = 1\pm 10\%$ requires larger  $|d_{\mu}|$. However,  the range of predicted $|d_{\mu}|$  is also growing, and at some point, for $k \lesssim 20$, it extends to $d_{\mu} = 0$. Further decreasing $k$ to about 2, even the  $R_{h\to \mu^+\mu^-} = 1 \pm 1\%$ range extends to $d_{\mu} = 0$. These findings are also clearly visible in Fig.~\ref{fig:edm_vs_k}, where we plot these contours of $R_{h\to \mu^+\mu^-}$ in the $k$ -- $|d_{\mu}|$ plane. 

\begin{figure}[t]
\includegraphics[width=0.6\linewidth]{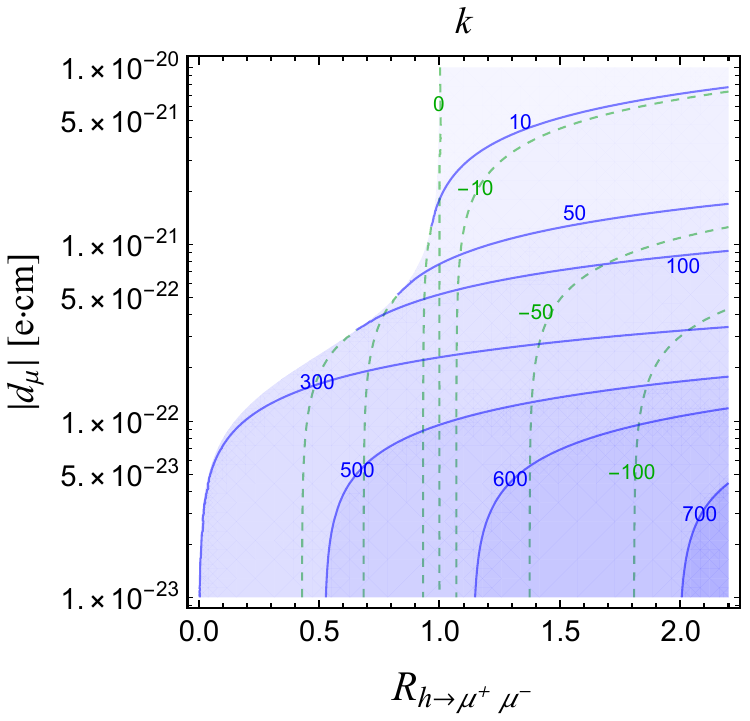} 
\caption{Contours of constant $k$ in the $R_{h\to\mu\mu}$ -- $|d_{\mu}|$ plane assuming the central value $\Delta a_{\mu}$.  The blue solid and green dashed curves indicate the larger and smaller solutions for $k$, respectively.}
\label{fig:k}
\end{figure}

Finally, allowing for any $R_{h\to \mu^+\mu^-}$ as a future possible measured value (up to the current limit), in Fig.~(\ref{fig:k}) we plot contours of $k$  in the $R_{h\to \mu^+\mu^-}$ -- $|d_{\mu}|$ plane assuming the central value $\Delta a_{\mu}$. 
Note that there are two values of $k$ for the same $R_{h\to \mu^+\mu^-}$ and $|d_{\mu}|$, corresponding to the two roots of Eq.~\ref{eq:ellipse}, except for the boundary of the shaded region. The boundary corresponds to the upper limit on possible $\mu$EDM if $R_{h\to \mu^+\mu^-} < 1$,
\begin{equation}
|d_{\mu}| \leq \frac{e \,\Delta a_{\mu}}{2m_\mu}\sqrt{\frac{R_{h\to \mu^+\mu^-}}{1-R_{h\to \mu^+\mu^-}}}.
\end{equation}
Note that models with small $k$ can generate $d_{\mu}$ up to the current experimental upper limit, $|d_{\mu}|=1.8\times 10^{-19}\;{\rm e\cdot cm}$~\cite{Muong-2:2008ebm}.

\noindent
{\it Discussion and conclusions.} 
We have seen that every model with chirally-enhanced contributions to  $\Delta a_{\mu}$ can be parametrized by the $k$ factor that relates the dipole operator to the contribution to the muon mass and thus specifies the correlation between $\Delta a_{\mu}$, $d_{\mu}$, and $R_{h\to \mu^+\mu^-}$. In the SM with VLs the $k$ factor is fully fixed by quantum numbers; in similar models with extended Higgs sectors a mixing parameter will enter $k$, for example $\tan\beta$ in the 2HDM; and in loop models $k$ is directly related to the coupling responsible for chiral enhancement. Through this correlation large classes of models or vast ranges of model parameters can be efficiently tested.

The $R_{h\to \mu^+\mu^-}$ is expected to be measured with $\sim10\%$ precision at the LHC and $\sim 1\%$ at the  hadron version of the Future Circular Collider~\cite{Abada:2019lih}. The limits on $\mu$EDM are expected to reach $|d_{\mu}|\sim 1\times 10^{-21}\,{\rm e\cdot cm}$ at the Muon $g-2$ experiment at Fermilab~\cite{Chislett:2016jau}, and could reach $6\times 10^{-23}\,{\rm e\cdot cm}$ at the Paul Scherrer Institute~\cite{Adelmann:2021udj}. Thus, near future measurements have the potential to reduce the number of SM extensions with vectorlike leptons to one specific $\mathcal{Q}$, or rule out all of them,  irrespectively of the scale of new physics or the size of couplings.
For 2HDM type-II, already the LHC measurement of $R_{h\to \mu^+\mu^-}$ will limit $\tan\beta$ to $\gtrsim 6$, and for loop models, it will limit the size of the coupling resulting in chiral enhancement, again irrespectively on other details of the model. This immediately sets the upper bound for the scale of new physics to $\sim 18$ TeV for the 2HDM~\cite{Dermisek:2020cod, Dermisek:2021ajd} and, for example, $\sim14$ TeV for FFS loop models with $SU(2)$ doublets and singlets and $\mathcal{Q} =1$. Improving the measurement of $R_{h\to \mu^+\mu^-}$ to within $1\%$ will further reduce these upper limits to $\sim 10$ TeV and $\sim 8$ TeV, respectively. Similar reasoning can be used to obtain an upper limit on the lightest new particle in any given scenario.
Thus, the correlation between $\Delta a_{\mu}$, $d_{\mu}$, and $R_{h\to \mu^+\mu^-}$ can test most efficiently the high end of the spectrum that is far beyond the reach of currently envisioned future colliders.

The discussion in this Letter has been limited to couplings necessary to generate a chirally-enhanced contribution to $\Delta a_{\mu}$ through an interaction of the SM Higgs to new particles. Effects of other possible dimension 6 operators involving muon fields, SM Higgs doublet and derivatives can be absorbed into the definition of the muon Yukawa coupling, $y_\mu$, or are parametrically suppressed by $m_\mu/v$, similar as in the discussion in Ref.~\cite{Dermisek:2021mhi}. While not a generic feature, in some models the dimension-5 operator $l_{L}l_{L}HH$ may also be generated which would be subject to additional constraints from the neutrino sector. However additional couplings in a given model, not contributing to $\Delta a_{\mu}$ might in principle enter the formula for $k$ (or even generate a complex $k$ parameter), for example scalar quartic couplings involving new scalars and the SM Higgs doublet~\cite{Thalapillil:2014kya,Crivellin:2021rbq}. In such cases $R_{h\to \mu^+\mu^-}$ still carves an ellipse in the plane of dipole moments but with the center shifted (to non-zero  $d_{\mu}$ for complex $k$) depending on the size of additional couplings. Furthermore, in certain models there can be sizable contributions to $\Delta a_{\mu}$ from other operators due to renormalization group mixing, for example from four-fermion operators in models with leptoquarks~\cite{Gherardi:2020det,Aebischer:2021uvt}. However, the couplings required also necessarily generate $C_{\mu H}$ with the same phase resulting in a shift of $k$ by a real number. These  effects, together with the general study of a complete set of dimension 6 operators, and other classes of models~\cite{Guedes:2022cfy} will be discussed elsewhere~\cite{Dermisek}.

\vspace{0.3cm}
\acknowledgments
The work of R.D. was supported in part by the U.S. Department of Energy under Award No. {DE}-SC0010120. TRIUMF receives federal funding via a contribution agreement with the National Research Council of Canada.

%
%
%

\vspace{0.05cm}





\end{document}